\documentclass[11pt, preprint]{aastex6}

\newcommand{\kms}{\mbox{${\rm km\, s^{-1}}$}}

\newcommand{\ascii}{{\sc ascii}}
\newcommand{\unix}{{\sc unix}}

\usepackage{graphicx,amsmath,amssymb,epsfig}
\usepackage{fancyvrb}
\usepackage[T1]{fontenc}
\usepackage{multirow}
\usepackage{color}

\shorttitle{VoigtFit}
\shortauthors{J.-K. Krogager}

\begin{document}


\title{{\large\bf VoigtFit}:\\ {\small A Python package for Voigt profile fitting}}

\author{Jens-Kristian Krogager\altaffilmark{1,2}}
\altaffiliation{E-mail:}

\altaffiltext{1}{Institut d'Astrophysique de Paris, CNRS-UPMC, UMR7095, 98bis bd Arago, 75014 Paris, France}
\altaffiltext{2}{Niels Bohr Institute, University of Copenhagen, Juliane Maries Vej 30, 2100 Copenhagen \O, Denmark}

\email{krogager@iap.fr}

\begin{abstract}

\noindent
I present a Python package developed for fitting Voigt profiles to absorption lines. The software fits multiple components for various atomic lines simultaneously allowing parameters to be tied and fixed. Moreover, the code is able to automatically fit a polynomial continuum model together with the line profiles. Lastly, a physical model can be used to constrain thermal and turbulent broadening of absorption lines. The code can be run with interactive features such as manual continuum placement locally around each line, manual masking of undesired fitting regions, and interactive definition of velocity components for various elements. This greatly improves the ease by which the initial guesses can be estimated. Since the code is written in pure Python, it can easily be scripted and modified to fit the user's needs. 
The code uses a $\chi^2$ minimization approach to find the best solution. The code and a set of test-data together with the full documentation is available on GitHub.

\end{abstract}


\section{Introduction}
Voigt profiles are commonly used to infer properties of absorption systems, e.g., intervening quasar absorbers or gamma-ray bursts. For this purpose, software has been developed to handle the fitting of such absorption profiles taking into account the line broadening by the instrument that was used to obtain the data. Probably, the two most widely used software packages are {\sc vpfit}\footnote{\url{https://www.ast.cam.ac.uk/~rfc/vpfit.html}} (Carswell \& Webb) and {\sc FitLyman} \citep{FitLyman}. Both have been used extensively in the literature and provide robust means of inferring the line properties: the column densities of the absorbing species, the accurate redshifts and the broadening parameters (also called the $b$-value).
While these software packages work well, they can be difficult to install on newer computers since they were designed for older architecture. Moreover, they are not easily scripted or included in larger projects.

I here present a new software package written entirely in Python. This code has the advantage that it is easily installed, and can be run both from a {\sc unix} shell and from a Python script. Moreover, the code is very interactive, allowing the user to define the number of components in a graphical interface (using matplotlib). The user can also define exclusion masks or fit a continuum to the region around the spectral lines.
Since the code is pure Python this gives the user a very powerful way to interact with the software in a fully scripted version. The code can therefore easily be incorporated in larger projects.

Recently, a Bayesian approach to absorption line fitting using Voigt profiles was developed \citep{BayesVP}. While this approach to Voigt profile fitting is very powerful and gives a good way of propagating errors, the Bayesian model becomes extremely complex in general cases with many components and lines fitted simultaneously in multiple spectra. Moreover, the implementation of tied parameters is not straight forward for complex line profiles.

In this document I will present the outline of how VoigtFit works, along with some details of the fitting procedure. The full documentation is available online on the GitHub\footnote{\url{https://github.com/jkrogager/VoigtFit}} page where it will be kept up to date regularly as the code develops.

\newpage

\section{Terms of Use}

\begin{center}
	Copyright \copyright\ 2013--2018 Jens-Kristian Krogager
\end{center}

This software is distributed under the open source MIT License.\\

Permission is hereby granted, free of charge, to any person obtaining a copy
of this software and associated documentation files (the "Software"), to deal
in the Software without restriction, including without limitation the rights
to use, copy, modify, merge, publish, distribute, sublicense, and/or sell
copies of the Software, and to permit persons to whom the Software is
furnished to do so, subject to the following conditions:\\

The above copyright notice and this permission notice shall be included in all
copies or substantial portions of the Software.\\

{\sc The software is provided ``as is'', without warranty of any kind,
express or implied, including but not limited to the warranties of
merchantability, fitness for a particular purpose and noninfringement.
In no event shall the authors or copyright holders be liable for any
claim, damages or other liability, whether in an action of contract,
tort or otherwise, arising from, out of or in connection with the
software or the use or other dealings in the software.}\\

\vspace{6mm}

\section{VoigtFit: Overview}
\label{overview}

The code fits different transitions from various ions of various elements simultaneously. In the following, the outline of the program will be presented with details about the steps of the needed input and computations. The program can be run in two ways; either directly from the \unix\ shell using an input parameter file, or by importing the VoigtFit module and running the program within a python script. Below we will first demonstrate the use of the input parameter file, which is the easiest way to get a quick start. The full documentation is available in the online \citet{manual}, where it will be kept up to date. The following is only an overview of the general structure of the program.\\

\subsection{Input and Setup}

The first and most crucial input is the spectral data to fit. This is given via the {\tt data} statement\footnote{A `statement' here refers to a programatic keyword in the parameter language. See more details in the online \citet{manual}.} in the input parameter file. The current version accepts either a FITS files with one (data only) or two (data and error) extensions or an \ascii\ table (white space or tab delimited) with at least two columns (wavelength and flux). Up to four columns can be used in the \ascii\ table format providing wavelength, flux, error and binary pixel mask\footnote{`1' for pixels to include in the fit and `0' for pixels to exclude.}, respectively.
If a FITS file is provided, then the FITS header must contain the keywords `CD1\_1' (or `CDELT1'), `CRVAL1', `CRPIX1' in order to recover the wavelength information. The flux array is assumed to be located in the first data extension. If an error array is present, this is assumed to be located in the second data extension.
Together with the input data file (or files) the corresponding spectral resolution for each spectrum should be given in units of \kms.
The wavelengths should be given in \AA, and are by default assumed to be corrected for atmospheric refraction (``air to vacuum'') and for heliocentric motion. If `air' wavelengths are given, the code can automatically convert these to vacuum wavelengths via the conversion from \citet{Edlen1953}. This is applied by including the {\tt air} keyword in the {\tt data} statement. For more details about the input parameter file format, see the online documentation.

The flux and error spectrum can either be given as observed flux or continuum normalized flux (specified through the {\tt norm} keyword).\\

The user then has to specify which transitions to fit following a general naming scheme of `ion' and rest-wavelength in \AA\ (with no digits) concatenated by an underscore; e.g., `SiII\_1526' or `HI\_1215' (a full list of available lines are visible in the atomic data file). In cases of conflicting wavelengths for the same ion (such as in densely spaced transitions of H$_2$ or fine-structure transitions of \ion{C}{1}), more digits are included in order to resolve the ambiguity.  For fine-structure levels of neutral carbon, the software knows the most common line-complexes and can automatically add the corresponding fine-structure lines for a single ground state transition.  The same is true for molecular transitions of CO and H$_2$, where the rotational levels of vibrational `bands' can be defined by referencing the vibrational and rotational levels one wants to fit.

The program automatically creates a fitting `region' around each line and if several fitting regions overlap, they are concatenated into one `region'. One `region' can therefore hold several transitions. In order for this automatic definition to work, the user must specify a systemic redshift (in future implementations, a first guess for the absorber redshift will automatically be generated using cross-correlation algorithms). For now, the size of the fitting region can be controlled by the `velocity span' parameter. By default, a symmetric cutout of $\pm$500~\kms\ around line-center will be used.\\

After the lines have been declared, the user must specify how many components (at least one) should be fitted for each ion, for which transitions have been defined in the step above. The components are defined by giving the redshift of the component (or velocity relative to the systemic redshift), the $b$ parameter, and the logarithm of the column density. Note that the parameters for different transitions of the same ionization state are fixed such that the relative velocity, $b$-parameter and column density for each component of \ion{Fe}{2}$\,\lambda2374$ will be identical to those of \ion{Fe}{2}$\,\lambda2344$, etc..

Moreover, the various parameters for the components can be tied to each other, e.g., the redshifts and $b$-parameters of two ions can be tied to each other, or column density ratios can be applied. This is useful when fitting heavy species of the same ionization state in damped Ly$\alpha$ Absorbers where the $b$-parameters are indistinguishable.

Components can also be specified using a graphical interface. When using the interactive interface, the user must select the location of the components by clicking in a graphical window. This will select the redshifts and a first estimate of the optical depth from the depth of the absorption profile. The column density is calculated by setting the initial $b$-parameter equal to the resolution element.
The interactive definition of components can be run several times for various lines of the same ion, in which case all the components will be used. This is helpful if the line structure is complex and several weak and strong lines are needed to constrain the number of components.

Lastly, components can easily be copied from one ion to another. This way the user can skip the tedious task of copying and pasting in the parameter file. The copying of components assumes that the components should be tied to one another, however, this assumption can be overruled.
If certain components turn out to be unconstrained for one ion, then these components can be removed. The code will not automatically assess whether or not a component is significant. In most cases, the fit will not converge properly and uncertainties of zero are returned. This is usually an indication that the user is trying to overfit the given data.\\

\subsection{Continuum Normalization}

After defining transitions and component structure, the program will normalize the data unless normalized data were passed as input. The continuum normalization can be done in an interactive window or automatically using Chebyshev polynomials. In case of interactive continuum fitting, the user will be prompted with a pop-up window for each fitting region defined above. The continuum can then be fitted as a linear interpolation between two ranges in the left and right parts of the spectral region. Alternatively, the user can specify a range of spline points which will be used for a smoother (yet less accurate) continuum estimate. This behaviour is controlled by the {\tt norm\_method} statement in the parameter file.

The easiest way to obtain a good continuum fit is to use the Chebyshev polynomials (by default up to 1$^{\rm st}$ order is included). This will define a continuum function of the form:
\begin{equation}
	C(\lambda) = \sum_{n=0}^{\tt C\_order} p_n T_n(\lambda)~,
\end{equation}
where $T_0(\lambda) = 1$, $T_1(\lambda) = \lambda$ and
$T_{n+1}(\lambda) = 2\lambda T_{n}(\lambda) - T_{n-1}(\lambda)$, and {\tt C\_order} is the highest order to include. The coefficients $p_n$ are free variables that will be fitted together with the line parameters. This generally gives the best results, however, this requires the user to mask out parts of the data that do not contain either continuum or line information. Contaminating lines will bias the continuum fit, which in turn biases the line fit. The highest order of polynomials to include is controlled by the {\tt cheb\_order} statement (also available as the alias {\tt C\_order}). In order to disable the use of Chebyshev polynomials, the user can provide a negative order, e.g., {\tt C\_order = -1}.\\

\subsection{Mask Definition}

The user can enable interactive line masking as an alternative to passing the mask in the \ascii\ table data. The code will then prompt the user to mask unwanted parts of the fitting regions in a graphical window. The {\it exclusion} masks are defined by clicking on the left and right boundaries of the part of the data that should be excluded. Several parts of data can be masked in one region. And the masking can be updated during various steps to add more masks until the user is happy with the mask.
In general, everything which does not solely contribute to either continuum or the lines that will be fitted should be masked out. Note that the mask is an {\it exclusion} mask and marks pixels that should {\it not} be used for the fit. Everything that is within a fitting region (and is not masked out) is included in the fit.

The interactive masking is by default turned off, however, it can be easily be applied for all lines or used only for certain, user-defined lines (see the online documentation for details).\\

\section{Optical Depth and $\chi^2$ Minimization}

The absorption line arising from a transition $i$ of element $X$ can be
described by the optical depth of the transition, $\tau$, which is determined by
the column density of the element $X$ along with a set of atomic parameters
describing the line strength, $f_i$, the damping constant, $\Gamma_i$,
and the resonance wavelength, $\lambda_i$, for the transition, $i$:

\begin{equation}
    \tau _{i, X}(\lambda) = K_i\ N_X\ a_i \ H[a_i, x(\lambda)]~,
\end{equation}
\noindent
where $K_i$ and $a_i$ are given by:

\begin{equation}
    K_i \equiv \frac {e^2\ \sqrt{\pi}\ f\ \lambda_i} {m_e\ c\ b}\ \hspace{1cm} \mathrm{and}\ \hspace{1cm} a_i \equiv \frac {\lambda_i\ \Gamma _i} {4 \pi b}\ ,
\end{equation}

\noindent
where $e$ is the elementary charge, $m_e$ is the electron mass, $c$ is the speed of light, and $b$ is the broadening parameter (i.e., the combination of turbulent and Doppler broadening).

The line profile for a single component is determined by the {\it Voigt--Hjerting function}, $H(a_i, x)$:
\begin{equation}
    H(a_i, x) \equiv \frac{a_i}{\pi} \int_{-\infty}^{+\infty} \frac{e^{-y^2}} {(x-y)^2 + a_i^2}\ {\rm d}y~,
\end{equation}
\noindent
where $x(\lambda) = (\lambda - \lambda_i)/\lambda_D$ is the rescaled wavelength, $y=v/b$ is the velocity of the absorbing atom in units of the broadening parameter, $b$, and $\lambda_D = b \lambda_i/c$ is the Doppler wavelength. The Voigt--Hjerting line profile arises from the convolution of the Doppler broadening from thermal and turbulent motion with the Lorentzian contribution from natural line broadening.

Since the Voigt--Hjerting function is very laborious to evaluate for every iteration in the fit, an analytical approximation by \citet{TepperGarcia2006} is used instead:
\begin{equation}
    H(a_i, x) \approx h - \frac{a_i}{ x^2 \sqrt{\pi}} \left[ h^2\ (4 x^4 + 7 x^2 + 4 + 1.5 x^{-2}) - 1.5 x^{-2} - 1 \right]~,
\end{equation}
\noindent
where $h=e^{-x^2}$ \citep[see also][]{TepperGarcia2007}.
The total optical depth in a region is calculated as the sum over all components for all transitions defined in the fitting region and the resulting transmittance is given as:
\begin{equation}
    I(\lambda) = e^{-\tau(\lambda)}~.
\end{equation}

\hspace{1cm}

After calculating the transmitted flux for a given region, this intrinsic profile is broadened by the instrumental profile which is assumed to be Gaussian with a FWHM equal to the given spectral resolution. For all practical purposes, the Gaussian approximation is working well, and no effort has been put into allowing a user-defined convolution kernel due to the complexities in the code. Furthermore, determining the actual instrumental function can be extremely challenging. Finally, the spectral resolution is assumed to be constant in velocity, i.e., the resolution element increases linearly with wavelength.
The convolution is performed using the Fast Fourier Transform as implemented in {\tt scipy.signal.fftconvolve}. For this purpose, the intrinsic profile is evaluated on a subsampled, logarithmically binned wavelength grid and after the convolution is done, the broadened profile is interpolated back onto the observed wavelength grid.\\

In case the user has activated the Chebyshev continuum model, the broadened profile for each region is multiplied by the Chebyshev continuum model, $C(\lambda)$, as described above. Otherwise the continuum model is set to unity.
The final model for a given region is thus given as:
\begin{equation}
	\mathcal{M}(\lambda) = C(\lambda) \times [I(\lambda) \ast LSF]~,
\end{equation}
where $LSF$ denotes the instrumental `line spread function' (here assumed to be Gaussian).

Lastly, the fit is performed using a $\chi^2$ minimization (default method is Levenberg--Marquardt) as implemented in the python package {\tt lmfit}.
The $\chi^2$ is calculated as the sum of residuals over all regions:

\begin{equation}
    \chi^2 = \sum_{k=0}^{\mathcal K} \sum_{\lambda} \left( \frac{\mathcal{F}_{k}(\lambda) - \mathcal{M}_{k}(\lambda)}{\sigma_{k}(\lambda)} \right)^2~,
\end{equation}

\noindent where $\mathcal{F}_{k}$ denotes the flux of the $k^{\rm th}$ fitting region, $\mathcal{M}_k$ denotes the model spectrum of the given region, and $\sigma_k$ refers to the uncertainty of the spectral data.

The user can specify any of the other vector minimization methods implemented in {\tt lmfit} via the input parameter file (e.g., {\tt fit-options method='nelder'} to use `Nelder--Mead' minimization). Similarly, the fit tolerance can also be altered, see details of the parameter file format in the online documentation.\\

\section{Output}

When the fit has converged, the best-fit parameters for each component of all fitted ions will be printed to screen as well as saved to a parameter file (`{\tt .fit}'). The various fitting regions, their best-fit profiles and the residuals will be shown in a graphical window as well as saved to a {\tt .pdf} file. The data and best-fit profile are furthermore saved to an \ascii\ file to allow the user to plot the fit results in any way that the user might like. Lastly, the whole data structure is saved as a `dataset' using the HDF5\footnote{\url{http://www.hdfgroup.org/HDF5/}} \citep{hdf5} data format. That way, the user can reload a given fit.
If the fit is re-run in the same directory, any previously saved dataset is automatically loaded such that masking and continuum normalization does not need to be performed again -- unless specified by the user.\\

\section{Example of Scripting}
\label{scripting}

As mentioned above, the code can be run either directly from the \unix\ shell with a parameter file or as part of a Python script. In this section, I will highlight how the Python interface works. The program was originally written with the scripting interface in mind, which is why this way of working with the code is much more powerful than the parameter file, which is somewhat limited to more standard cases.
The optimal way of using the program is to run it within IPython. This allows the user the maximum amount of freedom for modifying a dataset on the fly before fitting the absorption lines.

The backbone of the code is the `DataSet' class which handles most of the bookkeeping. The transitions to fit, the definition of atomic parameters in the `Line' class, the component structures, and the parameters are all managed by the `DataSet' class. This class is instantiated with a single input, namely the systemic redshift. Hereafter, the various elements of the steps described above can be added and modified by using the methods of the `DataSet'. In order to add spectral data (similar to the {\tt data} statement in the parameter file), the user must call the {\tt .add\_data} method which takes the input wavelength, flux, spectral resolution, error spectrum, and spectral mask as input. This allows the user to read in arbitrary data structures since the user can create a snippet of code for reading the data (possibly from several files\footnote{This is useful if the user has flux and error in different single extension FITS files.}) before passing it to the `DataSet'.

Hereafter, the user can add transitions (or lines), components, molecular bands, atomic fine-structure complexes (e.g., \ion{C}{1}$^{*}$ and \ion{C}{1}$^{**}$).

The full description of the interface is available and kept up to date online. Below, I will demonstrate some cases, where the scripting capabilities allow the user to fit a physically motivated model by imposing non-standard parameter constraints. For the full reference, see the example scripts in the folder `script' on GitHub\footnote{\url{https://github.com/jkrogager/VoigtFit}}.\\

\subsection{Thermal Broadening Model}
The physical model that we will implement below is motivated by the fact that different ions in a single, isothermal gas phase in thermal equilibrium governed by a temperature, $T$, and turbulent velocity, $b_{\rm turb}$, follow a distribution of broadening parameters, here on referred to as the effective $b$-parameter. This can be described as follows:

\begin{equation}
	b_{\rm eff}^2 = b_{\rm turb}^2 + \frac{2 k_B T}{m}~,
\end{equation}
where $k_B$ is Boltzmann's constant and $m$ is the atomic mass of the given ion. 

We can use this physically motivated information to reduce the number of parameters in cases where we have many transitions of varying mass available. The way to implement this in the script is by using the `expression' option of the fit parameters in {\tt lmfit}. These are set as algebraic expressions written in string format, using the parameter `names', e.g., the $b$-parameter of the first component of \ion{C}{2} would be called {\tt 'b0\_CII'} (notice the zero-indexing). We can therefore use such expressions to redefine our problem in terms of the various $b$-parameters as a problem in terms of $T$ and $b_{\rm turb}$. Note that this is a more accurate solution to simply tying all the $b$-parameters of all ions to have the same value; however, this model requires that many transitions for ions with a wide range of masses be available.

First, we add the new parameters to the `DataSet' parameter structure, through the attribute {\tt .pars}:

\begin{Verbatim}[xleftmargin=.1in]
	DataSet.pars.add_parameter('T_0', value=4000., vary=True, min=0.)
	DataSet.pars.add_parameter('turb_0', value=5., vary=True, min=0.)
\end{Verbatim}

Notice that we set a lower limit to avoid getting unphysical negative values. These parameters will be defined for each component following a similar naming scheme, such that the temperature of the first component will be called {\tt 'T\_0'}. Alternatively, if all components should follow the same temperature and turbulent broadening, then the user can just define one set of $T$ and $b_{\rm turb}$ parameters.

Secondly, the user must now link the physical parameters to the effective $b$-parameters defined for each component. This is done by accessing the {\tt expr} attribute of the parameters:

\begin{Verbatim}[xleftmargin=.1in]
	model_constraint = "sqrt(turb_0**2 + 2*kB*T_0/m)"
	DataSet.pars['b0_CII'].set(expr=model_constraint)
\end{Verbatim}

Here {\tt kB} refers to Boltzmann's constant and {\tt m} refers to the atomic mass of the ion in question. The atomic mass of a given ion is stored in the `Line' class for the given transition (this is indicated in the scripting example). It is easier to define the constants as variables and pass them to the string (see example script). Otherwise, the user must define these constants as fit parameters with {\tt vary=False}. The expressions should be defined for all components of all ions defined.

The program will now optimize the parameters $T$ and $b_{\rm turb}$ instead of the individual $b$-parameters for all ions. A synthetic spectrum is available to run the test script (see the `test\_data' directory on GitHub). The results of the test script are summarized in Table~\ref{tab:phys_model}.

The best-fit parameters are fairly well recovered in this case where the signal-to-noise is high (SNR$\sim$50); however, as the noise increases or fewer lines are available to constrain the fit (as is usually the case in real data), the fit becomes harder to constrain and may in many cases not reach convergence.\\

\begin{table}
	\caption{Comparison of input and output from physical model fit.
			 \label{tab:phys_model}}
	\begin{center}
		\begin{tabular}{ccc}
			\hline
			Parameter         &  Input  &  Best-fit \\
							  & (\kms)  &    (K)    \\[2pt]
			\hline
			$T_0$	          &   4370  &  3380$\pm$268 \\
			$T_1$	          &   8600  &  9280$\pm$206 \\[2pt]
			\hline
			$b_{\rm turb, 0}$ &   4.60  &  4.63$\pm$0.02 \\
			$b_{\rm turb, 1}$ &   5.20  &  5.16$\pm$0.02 \\[2pt]
			\hline
		\end{tabular}
	\end{center}
\end{table}

This thermal model can also be run from the parameter file using the keyword {\tt thermal model} (see details in the online \citet{manual}). However, the functionality is slightly reduced for the time being.\\

\subsection{Population of Molecular Levels}

Another case where a manual definition of parameter ties can be useful is when analysing low signal-to-noise data of molecular absorption in the rest-frame ultra-violet wavelength range. As an example, we will here look at the analysis of the CO molecule, whose various rotational levels are populated primarily by the thermal excitation of the cosmic microwave background \citep[e.g.,][]{Srianand2008a, Noterdaeme09b}. Under the assumption of an isothermal gas-phase, the relative column densities of the different rotational $J$-levels (and column densities, assuming that the physical density traces the column density) follow a Boltzmann distribution with a single temperature ($T_{\rm ex} \approx T_{\textsc{cmb}}$).

The fitting process can then be simplified significantly by tying the relative column densities of the higher rotational levels to the ground level by applying a calculated offset given by the Boltzmann distribution. Like in the example given above, this can easily be achieved by using VoigtFit in scripting mode:

\begin{Verbatim}[xleftmargin=.1in]
	dN1 = boltzmann_population(T_CMB, J=1)
	dN2 = boltzmann_population(T_CMB, J=2)

	model_constraint = "logN0_COJ0 + %f" % dN1
	DataSet.pars['logN0_COJ1'].set(expr=model_constraint)

	model_constraint = "logN0_COJ0 + %f" % dN2
	DataSet.pars['logN0_COJ2'].set(expr=model_constraint)
\end{Verbatim}

Here the function {\tt boltzmann\_population} should define the relative column densities for a given $J$-level with respect to the ground state (in logarithmic units). The module {\tt VoigtFit.molecules} constains functions to calculate these ratios. The offset in $\log(N)$ can then be added to the ground level column density by setting an algebraic relation between the two parameters as we saw in the previous example. If more components are present for each $J$-level, then similar expressions should be defined for each component, i.e., parameter names {\tt 'logN1\_COJ0'}, {\tt 'logN2\_COJ0'}, etc..\\

\subsection{Partial Coverage Effects}

In cases where the background source is more extended than the absorbing medium, a residual flux will leak through to the observer effectively diluting the absorption line. This is especially evident for saturated lines, where the saturated core of the line no longer reaches zero, but instead reaches a non-zero level depending on the fraction of the background source covered by the absorbing medium. The amount of unabsorbed flux is called the `line flux residual`, $f$, which relates to the covering fraction as $C_f = 1 - f$. This effect is observed in a handful of cases of cold gas absorbers where the absorption lines fall on top of broad emission lines from the background quasar. Since the emission line region for such broad lines is much more extended than the continuum source, the transitions from the compact cold gas phase (typically molecular hydrogen or neutral species like \ion{C}{1}) falling on top of the broad emission lines will have a residual flux since part of the broad emission line flux reaches the observer unabsorbed.
For examples and detailed interpretations of this phenomenon, see \citet{Balashev2011} or \citet{Bergeron2017}. The results of using VoigtFit to fit partial coverage for \ion{C}{1} and its fine-structure lines is demonstrated in the analysis of \citet{Krogager2016a}.

At the time of writing, this functionality is only available through scripting. In future versions of the code, the ability to fit specific lines with a variable line-flux residual will be added.

The way to script this is by accessing the fitting regions of the lines that are affected by partial coverage and modify that given fit region. In the overview below, the line flux residual, $f$, is fitted for the \ion{C}{1}$\lambda1656$ line. Since the synthetic profile calculation cannot be altered directly (without changing the source code), the way to include $f$ is to modify the observed data to recover the unaffected optical depth. In a simple case where the absorbing medium covers all of the continuum source and only a part of the broad emission line flux, the resulting observed, normalized flux will be:

\begin{equation}
	F_{\rm norm}(\lambda) = (1-f)e^{-\tau(\lambda)} + f~,
\end{equation}

\noindent
where $\tau$ refers to the optical depth. We can therefore reformulate the problem to recover the optical depth which we can easily model using VoigtFit:

\begin{equation}
	e^{-\tau(\lambda)} = \frac{F_{\rm norm}(\lambda) - f} {1-f}~.
\end{equation}

\noindent
By rescaling the observed, normalized spectrum in the given fitting region, we can include the effect of partial coverage. The way to implement this in a Python script would be as follows:

\begin{Verbatim}
	import VoigtFit
	
	dataset = VoigtFit.DataSet(z_sys)
	
	# -- Load Data and feed to the dataset
	# Add lines here to read the data in whatever format you have
	dataset.add_data(wl, flux, res, err=err,
			 mask=mask, normalized=True)
	
	# -- Define lines and components
	# We keep it simple for now
	dataset.add_fine_lines('CI_1560')
	dataset.add_fine_lines('CI_1656')
	dataset.add_component_velocity('CI', 0., 5., 14.)
	dataset.copy_components(from_ion='CI', to_ion='CIa')
	dataset.copy_components(from_ion='CI', to_ion='CIb')
	
	# -- Define the line flux residual
	f = 0.15
	
	# -- Find the fit region that contains CI_1656
	region, = dataset.find_line('CI_1656')
	flux = region.flux
	
	# -- Update the data taking the LFR into account
	flux_new = (flux - f)/(1 - f)
	region.flux = flux_new
	region.err /= (1 - f)
	
	# -- And fit the dataset
	dataset.prepare_dataset()
	popt, chi2 = dataset.fit()
	
\end{Verbatim}

Notice that the result from {\tt .find\_line()} is a list\footnote{The reason for this is that the code can fit several overlapping spectra at once, hence one line can be defined in several regions.}, but since in this case I am sure there is only one \ion{C}{1} line defined, I can safely unpack it into a single variable.

This only applies one single value of $f$ to the fit, so the user will have to loop over a grid of $f$-values and store the $\chi^2$ value to calculate the optimal value of $f$. As mentioned above, this functionality will be improved in future versions of the code.\\

\section{Application on real data}
\label{test}

\begin{figure}
	\centering
	\includegraphics[width=0.8\textwidth]{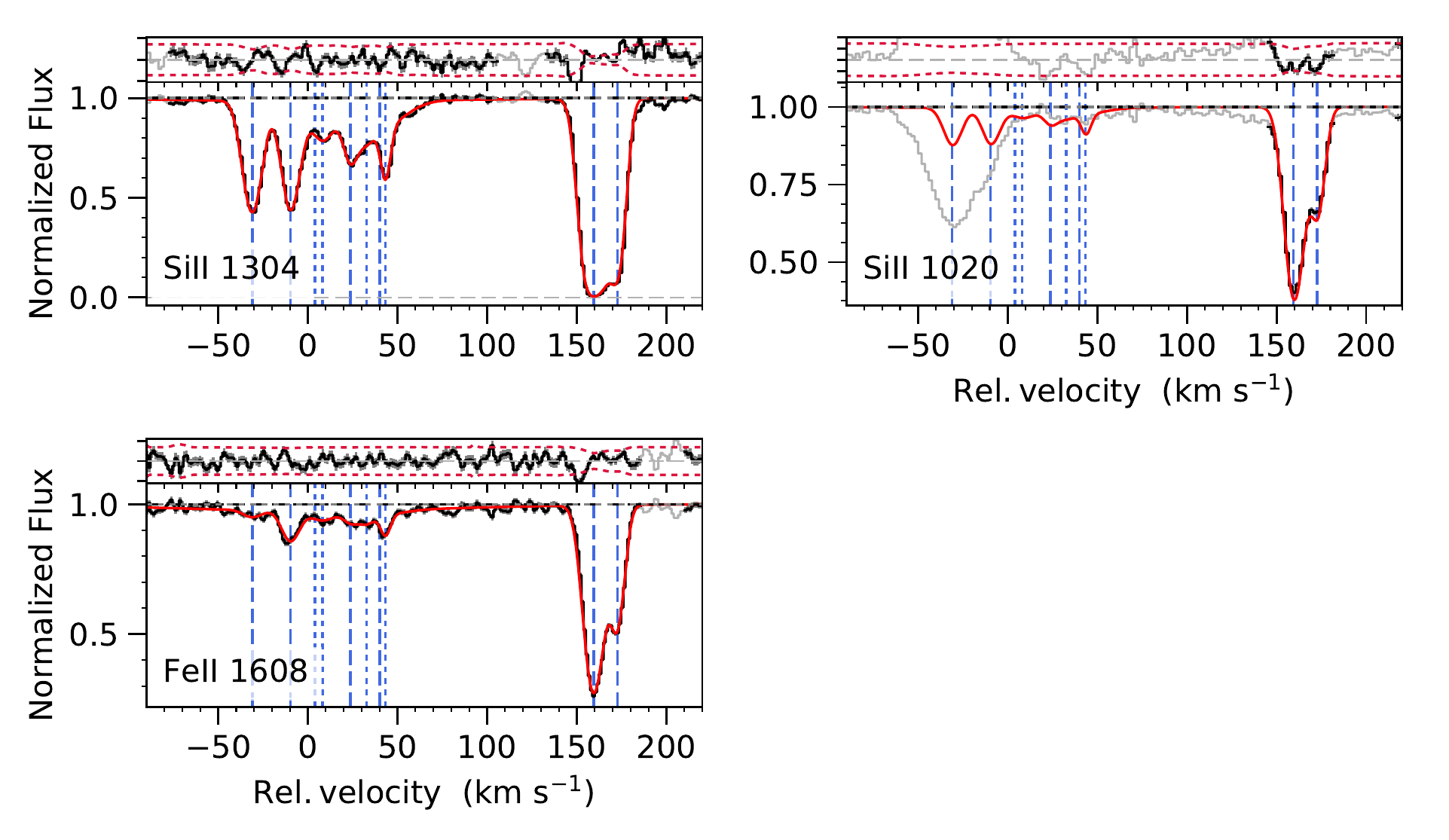}
	\includegraphics[width=0.8\textwidth]{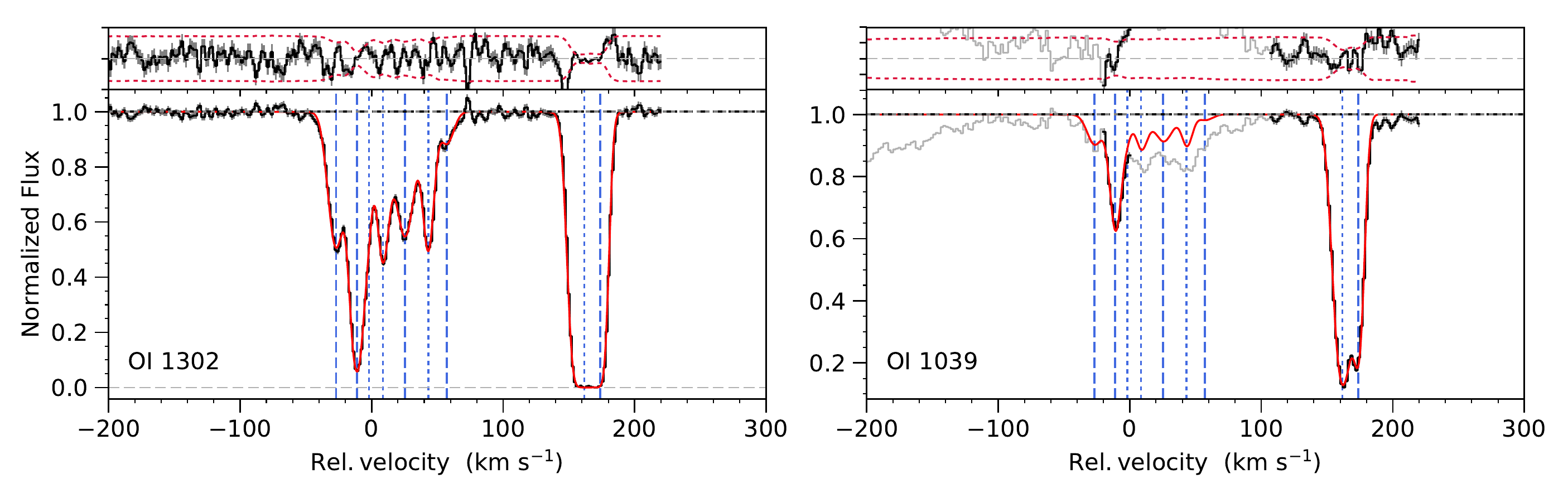}
	\caption{Best-fit profiles for fitted absorption lines of the $z_{\rm sys}=3.08815$ absorber towards the quasar Q0420$-$388. The gray parts of the spectra have been masked out and were not included in the fit due to blending with contaminating lines. The red, solid lines show the best-fit profiles, and the dashed vertical lines mark the positions of each component. The top part of each panel shows the residuals in each region, where the dotted red lines show the 3$\sigma$ error spectrum.}
	\label{fig:Q0420}
\end{figure}

As a test to showcase the capabilities of the program, I have re-analysed high resolution data of the quasar Q0420$-$388 from the UVES spectrograph mounted at the Very Large Telescope at Paranal Observatory, Chile.
The absorption line system at $z=3.09$ shows strong neutral oxygen lines together with strong singly ionized lines of silicon and iron (among others). In Fig.~\ref{fig:Q0420}, I show the results of a 10-component fit to some of the \ion{Si}{2} and \ion{Fe}{2} lines and a 9-component fit to the \ion{O}{1} lines.
The best-fit total column densities are $\log\, N({\rm Fe\, \textsc{ii}}) = 14.09 \pm 0.02$, 
$\log\, N({\rm Si\, \textsc{ii}}) = 14.65 \pm 0.01$, and
$\log\, N({\rm O\, \textsc{i}}) = 15.34 \pm 0.01$. The system has been analysed previously by \citet{Carswell1996} using more noisy and lower resolution data (ranging from 8 to 13~km~s$^{-1}$), however, the obtained values are consistent. \citet{Carswell1996} find $\log\, N({\rm Fe\, \textsc{ii}}) = 13.95 \pm 0.10$, 
$\log\, N({\rm Si\, \textsc{ii}}) = 14.47 \pm 0.09$, and $\log\, N({\rm O\, \textsc{i}}) = 15.37 \pm 0.04$.
The slight discrepancies for \ion{Si}{2} and \ion{Fe}{2} might stem from the fact that I fit the lines with tied broadening parameters for all components, whereas \citet{Carswell1996} assume a fixed temperature (of $T=10^4$~K), and thus have slightly different $b$-values for Si and Fe. Moreover, the dominant component is a blend of two very strong and saturated lines, which may lead to differences between the two data sets due to the lower resolution and worse signal-to-noise in the previous analysis.

%

Other examples of the code can be seen in the recent works by \citet{Krogager2016a}, \citet{Krogager2017}, and Heintz et al. (2018) in review.

\newpage

\section{Summary}

I have here presented a Python package to perform Voigt-profile fitting of absorption lines. The code can be run from terminal using a parameter file or from a Python script. The scripting capabilities makes this package very powerful as it can easily be tweaked to each specific case. Moreover, the parameter file allows the user to quickly fit a system using interactive tools to define components, mask regions and normalize data.
The code is available online on GitHub or via Python package distribution tool {\tt pip}. The latest versions will be available on GitHub, and only larger releases of updates will be uploaded to the pip distribution service.
Similarly the \citet{manual} and the Python interface are described and kept up to date on the GitHub page (github.com/jkrogager/VoigtFit).
The software is free and distributed for the community to use, however, no warranty is provided.

Please report any bugs or suggestions for improvements via GitHub or by e-mail.\\

Lastly, if you make use of the software, please cite this paper.\\

\acknowledgements

{\sc Acknowledgements}\\

{\small
I want to thank J. Selsing, K. Heintz, J. Fynbo, C. Th{\"o}ne, L. Izzo, A. de Ugarte-Postigo, Bo Milvang-Jensen, and L. Christensen for their help in extensively testing the software and for their constructive feedback and ideas for new features. I also want to thank S. Balashev for helpful and constructive discussions.
Part of this work has been carried out with financial support from the Danish Council for Independent Research (EU-FP7 under the Marie-Curie grant agreement no. 600207) with reference DFF-MOBILEX--5051-00115. This work was initiated during my PhD studentship at the European Southern Observatory in Santiago, Chile under supervision by C\'edric Ledoux ({\sc Dec} 2012 -- {\sc Nov} 2014).}

\bibliographystyle{apj}

\end{document}